\RequirePackage{fix-cm}
\documentclass[twocolumn,epjc3]{svjour3}
\smartqed  
\RequirePackage{graphicx}
\RequirePackage{amsmath} \RequirePackage{amssymb}
\usepackage{slashed}
\usepackage{bbm}
\usepackage{color}

\begin{document}

\title{Cold magnetized quark matter at finite density in a nonlocal chiral quark model}

\author{S.A. Ferraris\thanksref{addr1}
        \and
        D. G\'omez Dumm\thanksref{addr2,addr3} \and
        A.G. Grunfeld\thanksref{e1,addr1,addr2} \and
        N.N. Scoccola\thanksref{addr1,addr2}
}

\thankstext{e1}{e-mail: ag.grunfeld@gmail.com}

\institute{Departamento de F\'\i sica, Comisi\'on Nacional de
Energ\'{\i}a At\'omica, Av. Libertador 8250, (1429) Buenos Aires,
Argentina \label{addr1}
           \and
           CONICET, Godoy Cruz 2290, Buenos Aires, Argentina \label{addr2}
           \and
          IFLP, CONICET - Departamento de F\'{\i}sica,
Facultad de Ciencias Exactas, Universidad Nacional de La Plata, C.C. 67, 1900 La Plata, Argentina \label{addr3}
}

\date{Received: date / Accepted: date}

\maketitle

\begin{abstract}
We study the behavior of two-flavor dense quark matter under the
influence of an external magnetic field in the framework of a
nonlocal chiral quark model with separable interactions. The
nonlocality is incorporated in the model by using a Gaussian form
factor. It is found that for low and moderate
values of magnetic field there is a decrease of the
critical chiral restoration chemical potential $\mu_c$,
i.e.\ an inverse magnetic catalysis effect is observed. For larger values of
$eB$ the behavior of $\mu_c$ becomes more or less flat, depending
on the parametrization. Within the considered
parametrization range we do not find a significant growth of the
critical chemical potential for large magnetic fields, as occurs
in the case of the local NJL model.
\end{abstract}

\section{Introduction}
\label{intro} The phase structure of strongly interacting matter
under the influence of an external and uniform magnetic field
plays an important role in several physical scenarios, such as
heavy ion collisions~\cite{HIC}, cosmology~\cite{cosmo} and compact
star physics~\cite{duncan}. For this reason, the analysis of the
features of the QCD phase diagram in the presence of an
external magnetic field at both finite temperature and/or chemical potential has become an issue of great interest in
recent years~\cite{Kharzeev:2012ph,Andersen:2014xxa,Miransky:2015ava}. From
the theoretical point of view this kind of study requires to deal with QCD in
a low energy regime, where quantitative calculations are
extraordinarily difficult in view of the strong coupling involved. One way to
overcome this problem is by relying on lattice QCD (lQCD)
simulations; however, at finite density they present the harmful
sign problem. Then, to circumvent this obstacle, effective
models come into the picture as a powerful tool to describe the strongly
interacting matter phase diagram at finite densities. It is
crucial that effective models show consistency with lQCD
simulations at vanishing baryonic chemical potential, before
performing extrapolations to higher densities. One of the most
widely used approaches in this sense is the Nambu Jona-Lasinio (NJL) model~\cite{Nambu:1961tp,Nambu:1961fr}.
In this nonrenormalizable effective chiral quark model quarks interact through local four-point
vertices, leading to a spontaneous chiral symmetry breaking mechanism~\cite{Vogl:1991qt,Klevansky:1992qe,Hatsuda:1994pi}. As an improvement over local models, chiral quark models that include
nonlocal separable interactions have also been
considered~\cite{Schmidt:1994di,Burden:1996nh,Bowler:1994ir,Ripka:1997zb}. 
Since these approaches can be viewed as nonlocal extensions of the
NJL model, here we denote them generically as ``nonlocal NJL''
(nlNJL) models. It has been shown that nonlocal approaches include several advantages with respect to the local NJL scheme. For example, the inclusion
of smooth nonlocal form factors prevents ultraviolet
divergences in momentum space, and the nonlocality leads to a
momentum dependence in quark propagators that is consistent with LQCD
simulations. In addition, nlNJL models provide
a satisfactory description of the hadron properties at zero
temperature and density. A recent review with the description and applications
of nlNJL models to the analysis of the strong interaction matter
under extreme conditions can be found in Ref.~\cite{Dumm:2021vop}.

For magnetized quark matter, at zero temperature and vanishing
baryonic chemical potential, the results of low-energy effective
models of QCD as well as those of lQCD calculations indicate that {the size
of} light quark-antiquark condensates should {get increased with}
the magnetic field. Thus, the external field appears to favor the
breakdown of chiral symmetry, which is usually known as ``magnetic
catalysis''(MC) ~\cite{Gusynin:1994re}. On the contrary, close to
the chiral restoration temperature, lQCD calculations carried out
with realistic quark masses~\cite{Bali:2011qj,Bali:2012zg} show
that the condensates behave as nonmonotonic functions of $B$, and
this leads to a decrease in the transition temperature when the
magnetic field is increased. This effect is known as ``inverse
magnetic catalysis'' (IMC). In addition, lQCD calculations predict
an entanglement between the chiral restoration and deconfinement
critical temperatures~\cite{Bali:2011qj}. The observation of IMC
has become a challenge for effective
models~\cite{Andersen:2014xxa,Miransky:2015ava}. Indeed, most
naive effective approaches to low energy QCD predict that the
chiral transition temperature should grow with $B$, i.e., they do
not find IMC. Interestingly, the corresponding studies carried out
in the context of nlNJL
models~\cite{Pagura:2016pwr,GomezDumm:2017iex} show that the latter
are naturally able to describe, at the mean field level, not only
the IMC effect but also the entanglement between chiral
restoration and deconfinement transition temperatures. Moreover,
it is found that the behavior of the mass and decay constant of
the $\pi^0$ meson as functions of the external magnetic field are
also in agreement with lQCD
results~\cite{GomezDumm:2017jij,Dumm:2020muy}.

In the present work we study the behavior of cold strongly
interacting matter under a uniform, static magnetic field in the
framework of nlNJL models at finite density. Namely, our aim is to extend
the previous works~\cite{Pagura:2016pwr,GomezDumm:2017iex} to the finite
density region. This represents a further step towards
the analysis, within these models, of the effect of a strong magnetic field on the
full phase diagram of strongly interacting matter. It is important to
mention here that in the framework of the local NJL model several studies
of the behavior of cold magnetized strongly interacting matter
at finite chemical potential
have been reported in the literature
(see e.g.\ Refs.~\cite{Ebert:1999ht,Ebert:2003yk,Inagaki:2003yi,Menezes:2008qt,Boomsma:2009yk,Ferrari:2012yw,Ferrer:2012wa,Allen:2013lda,Avancini:2012ee,Grunfeld:2014qfa,Allen:2015qxa}).
One common feature found in those analyses is that for not so strong
magnetic fields the critical chemical potential for chiral symmetry restoration decreases
with increasing values of $B$. This phenomenon has been also dubbed
``inverse magnetic catalysis'' in Refs.~\cite{Preis:2010cq,Preis:2012fh},
where the same effect has been found in the context of the Sakai-Sugimoto
model. To avoid confusion with the IMC effect at finite temperature mentioned
above, here it will be referred to as ``inverse magnetic catalysis at finite
chemical potential'' ($\mu$IMC).
It is worth mentioning that the same effect is obtained when one considers quark-quark channels
that give rise to two flavour color superconducting quark matter~\cite{Fayazbakhsh:2010gc,Mandal:2012fq,Allen:2015paa,Mandal:2017ihr,Coppola:2017edn}.

The paper is organized as follows. In Sec.~2 we describe the
theoretical formalism of cold magnetized quark matter within 
nlNJL models at finite density. In Sec.~3 we show our results for
the chiral limit case and for finite current quark masses.
Finally, in Sec.~4 we sketch our conclusions.

\section{Theoretical formalism}

We start by quoting the Euclidean action for the nonlocal chiral
quark model under consideration. In the case of two light flavors
one has
\begin{eqnarray}
S_{E} & = & \int d^{4}x\ \Big[\bar{\psi}\left(x\right)
\left(-i\slashed{\partial} + \hat m \right) \psi\left(x\right)
\nonumber \\
& & \qquad \qquad -\, \frac{G}{2}j_{a}\left(x\right)j_{a}\left(x\right) \Big]\ ,
\label{EucAct}
\end{eqnarray}
where $\psi$ stands for the $u$, $d$ quark field doublet and
$\hat m =\mbox{diag}(m_u,m_d)$ is the current quark mass matrix.
Throughout this article we will work
in the isospin limit, thus we assume $m_u=m_d$. The currents $j_{a}(x)$ are given by
\begin{equation}
j_{a}\left(x\right) \ = \ \int d^{4}z\ \mathcal{G}(z)\,\bar{\psi}
\big(x + \frac{z}{2}\big)\Gamma_{a}\,\psi\big(x -
\frac{z}{2}\big)\ , \label{currents}
\end{equation}
where we have defined
$\Gamma_{a}=(\mathbbm{1},i\gamma_{5}\vec{\tau})$, and
$\mathcal{G}(z)$ is a nonlocal form factor that characterizes the
effective interaction.

To describe the behaviour of 
magnetized quark matter, we include in the effective action a
coupling to an external electromagnetic gauge field
$\mathcal{A}_{\mu}$. For a local theory, this can be done by
introducing a covariant derivative in the kinetic term of the
action in Eq.~(\ref{EucAct}), i.e.\ by changing
\begin{equation}
\partial_{\mu}\ \rightarrow\ D_{\mu}\equiv\partial_{\mu}-i\hat{Q}\mathcal{A}_{\mu}\left(x\right)\ ,
\end{equation}
where $\hat{Q} =\mbox{diag}\left(q_{u},q_{d}\right)$ is the
electromagnetic quark charge operator ($q_{u}=2e/3$,
$q_{d}=-e/3$). In the case of the nonlocal model studied here,
this replacement has to be supplemented by a contribution arising
from the nonlocal currents in Eq. (\ref{currents}). One
has~\cite{Ripka:1997zb}
\begin{eqnarray}
\psi\left(x-z/2\right) & \ \rightarrow\  & \mathcal{W}\left(x,x-z/2\right)\,\psi\left(x-z/2\right)\ ,
\end{eqnarray}
where the function $\mathcal{W}\left(s,t\right)$ is defined by
\begin{equation}
\mathcal{W}\left(s,t\right)\ =\ P\
\exp\left[-i\int_{s}^{t}dr_{\mu}\,
\hat{Q}\,\mathcal{A}_{\mu}\left(r\right) \right]\ .
\label{intpath}
\end{equation}
Here $P$ stands for path ordering, and $r$ runs over an arbitrary
path connecting $s$ with $t$. As it is usually done, the latter is
taken to be a straight line~\cite{Bloch:1952qkt}. For simplicity,
we restrict to the particular case of a constant and homogeneous
magnetic field, which is chosen to be orientated along the 3-axis.
To perform the analytical calculations we use the Landau gauge, in
which one has $\mathcal{A}_\mu = B\, x_1\, \delta_{\mu 2}$. With
this gauge choice the function ${\cal W}(s,t)$ in
Eq.~(\ref{intpath}) is given by
\begin{equation}
{\cal W}(s,t) \ = \ \exp\left[-\frac{i}{2}\,\hat
Q\,B\,(s_1+t_1)\,(t_2-s_2) \right] \ . \label{wstshwinger}
\end{equation}

To proceed, we perform a standard bosonization of the theory,
introducing scalar and pseudoscalar fields $\sigma(x)$ and
$\vec{\pi}(x)$, and integrating out the fermion fields. Moreover,
we consider the mean field approximation (MFA), assuming that the
field $\sigma(x)$ has a nontrivial translational invariant mean
field value $\bar{\sigma}$, while the mean field values of
pseudoscalar fields are zero. In the presence of the external
magnetic field, it is convenient to write the effective action in
a basis of Ritus functions \cite{Ritus:1978cj}. Details of this procedure can
be found e.g.\ in Refs.~\cite{Pagura:2016pwr,GomezDumm:2017iex}.

Since we are interested in the study of dense quark matter, we
consider a system at nonzero quark chemical potential $\mu$ ($\mu
= \mu_B/3$, where $\mu_B$ is the baryon chemical potential). Then,
the grand canonical thermodynamic potential can be obtained from
the effective action, including the chemical potential through the
replacement $\partial_4 \rightarrow \partial_4 - i \mu$ in the
kinetic term. In addition, to obtain the appropriate conserved
currents, this has to be supplemented by a modification of the
nonlocal currents in
Eq.~(\ref{currents})~\cite{GomezDumm:2017iex}. In practice, if the
Fourier transform of the nonlocal form factor $\mathcal{G}(z)$ is
denoted by $g(p)$, the latter has to be modified by changing $p_4
\rightarrow p_4  + i\mu$.  In this way, the thermodynamic
potential in the MFA is found to be given by
\begin{eqnarray}
\Omega_{\mu,B}^{\rm MFA} &=& \frac{\bar{\sigma}{^{2}}}{2G} -
\sum_{f=u,d} \frac{3 B_f}{2\pi}\int \frac{d^{2}p_\parallel}{(2\pi)^{2}} \nonumber\\
& \times& \left[\ln\left(p_\parallel^{2} +
{M_{0,p_\parallel}^{s_{f},f}}^{\,2\,}\right) + \sum_{k=1}^{\infty}
\ln \Delta_{k,p_\parallel}^{f} \right]\ , \label{gpotencialB}
\end{eqnarray}
where
\begin{eqnarray}
\Delta_{,p_\parallel}^{f} &=& \left(2k B_f + p_\parallel^{2} +
M_{k,p_\parallel}^{+,f} M_{k,p_\parallel}^{-,f}\right)^{2}
\nonumber \\ && +\, p_\parallel^{2}\left(M_{k,p_\parallel}^{+,f} -
M_{k,p_\parallel}^{-,f}\right)^{2}\ , \label{deltaKP}
\end{eqnarray}
with
\begin{equation}
M_{k,p_\parallel}^{\lambda,f} =
\left(1-\delta_{k_{\lambda},-1}\right) m_{c} + \bar{\sigma}\
g_{k,p_\parallel}^{\lambda,f}
\end{equation}
and
\begin{eqnarray}
g_{k,p_\parallel}^{\lambda,f}&=&\frac{4\pi}{|q_{f}B|}\left(-1\right)^{k_{\lambda}}\int \frac{d^{2}p_{\bot}}{\left(2\pi\right)^{2}}\ g\left(p_{\bot}^{2} + p_\parallel^{2}\right) \nonumber \\
&& \times\,
\exp\left(-p_{\bot}^{2}/B_f\right)L_{k_{\lambda}}\left(2p_{\bot}^{2}/B_f\right)\
. \label{funcg}
\end{eqnarray}
Here, we use the definitions $m_c = m_u = m_d$, $p_{\bot}=\left(p_{1},p_{2}\right)$,
$p_\parallel=(p_{3},p_{4}+i\mu)$ and $k_{\pm}=k-1/2\pm s_f/2$,
where $s_f=\mbox{sign}\left(q_{f}B\right)$. In addition, we denote
$B_f = \left|q_{f}B\right|$, while $L_{m}\left(x\right)$ are
Laguerre polynomials, with the usual convention
$L_{-1}\left(x\right)=0$. It can be seen that the functions
$M_{k,p_\parallel}^{\pm,f}$ play the role of constituent quark
masses in the presence of the external magnetic field. The index
$k$ is a quantum number that labels the so-called Landau energy
levels of quark fields.

In general, the expression in Eq.~(\ref{gpotencialB}) turns out to
be divergent. It can be regularized following the prescription in
Ref.~\cite{GomezDumm:2004sr}, namely
\begin{equation}
\Omega_{\mu,B}^{\rm MFA,reg} \ = \ \Omega_{\mu,B}^{\rm MFA} -
\Omega_{{\mu,B}}^{\rm free} + \Omega_{{\mu,B}}^{\rm free,reg}\ ,
\label{omereg}
\end{equation}
where $\Omega_{\mu,B}^{\rm free}$ is a ``free'' piece obtained
from $\Omega_{\mu,B}^{\rm MFA}$ by taking $\bar \sigma=0$ (and
keeping the chemical potential and the interaction with the
magnetic field). The regularized form of this free piece is given
by
\begin{eqnarray}
\Omega_{\mu,B}^{\rm free,reg} &=&
-\frac{N_{c}}{2\pi^{2}}\sum_{f=u,d}
\Big[ B_f^2 \ t_f  \nonumber \\
&& + \sum_k \theta\left(\mu-S_{kf}\right)\, \alpha_{k}\, B_f \,
v_{kf} \Big]\ , \label{omegafree}
\end{eqnarray}
where
\begin{equation}
t_f = \zeta^{'}\left(-1,x_{f}\right)  + \frac{x_{f}^{2}}{4}
-\frac{1}{2}\left(x_{f}^{2} - x_{f}\right)\ln  x_{f}
\end{equation}
and
\begin{equation}
v_{kf} = \mu\sqrt{\mu^{2}-S_{kf}^{2}} - S_{kf}^{2} \ \ln
\left[\frac{\mu + \sqrt{\mu^{2}-S_{kf}^{2}} }{S_{kf}}\right]\ .
\end{equation}
Here we denote $x_{f}=m_{c}^{2}/(2B_f)$, $\alpha_{k}=1 -
\delta_{0k}/2$ and $S_{kf}=\left(m_{c}^{2} + 2 k
B_f\right)^{1/2}$. In addition, we use the notation
$\zeta^{'}\left(z,x_{f}\right)=d\zeta\left(z,x_{f}\right)/dz$,
$\zeta\left(z,x_{f}\right)$ being the Hurwitz zeta function.

Given the regularized form of the mean field thermodynamic
potential, the mean field value $\bar{\sigma}$ can be obtained by
solving the gap equation
\begin{equation}
\frac{\partial\Omega_{\mu,B}^{\rm MFA,reg}}{\partial\bar{\sigma}}
\ = \ 0\ . \label{gapeq}
\end{equation}
In general, for each value of $\mu$ and $B$ several solutions of
this equation may exist. The most stable solution will be the one
corresponding to the absolute minimum of $\Omega_{\mu,B}^{\rm
MFA,reg}$. Once $\bar\sigma$ has been determined, the chiral quark
condensates $\langle \bar q q\rangle$, with $q =u,d$, can be
calculated from
\begin{equation}
\langle \bar q q\rangle \ = \ \frac{\partial \Omega_{\mu,B}^{\rm
MFA,reg}}{\partial m_q}\ .
\end{equation}

\section{Numerical results}

To obtain the numerical predictions that follow from the formalism
described in the preceeding section, it is necessary to specify
the particular shape of the nonlocal form factor. In the present
work we use a Gaussian function
\begin{equation}
g\left(p^{2}\right) \ =\ \exp\left(-p^{2}/\Lambda^{2}\right)\ ,
\end{equation}
where $\Lambda$ is a parameter that indicates the range of the
quark level interaction in momentum space. With this particular
choice of the form factor, the integral in Eq.~(\ref{funcg}) can
be performed analytically. One gets in this way
\begin{equation}
g_{k,p_\parallel}^{\lambda,f} =
\frac{\left(1-B_f/\Lambda^{2}\right)^{k_{\lambda}}}{\left(1+B_f/\Lambda^{2}\right)^{k_{\lambda}+1}}\
\exp\left(-p_\parallel^{2}/\Lambda^{2}\right)\ . \label{mass.eq13}
\end{equation}
Therefore, the model includes three free parameters, viz.\ the
effective momentum scale $\Lambda$, and the constants $m_c$
(current quark mass) and $G$ (coupling constant) that appear in
the effective action. In this work we consider
two situations. The first one, discussed in Sec.~3.1, corresponds
to the chiral limit, in which we fix $m_c=0$. Although not really
physical, this situation allows for a more clear identification of
the possible existing phases and it is, thus, quite useful from a
theoretical point of view. The second situation, analyzed in
Sec.~3.2, corresponds to the more realistic case in which a finite
current quark mass is considered. The way in which the model
parameters are determined in each case is discussed in the
corresponding subsections.

\subsection{Chiral limit}

In the chiral limit the number of model parameters reduces to only
two, viz.~$\Lambda$ and $G$. We consider a pa\-ram\-e\-triza\-tion
in which these are chosen so as to lead to a typical value for the
pion decay constant in the chiral limit, $f_{\pi,{\rm ch}}=90$
MeV, and a given value of chiral quark condensate at zero $B$ and
$\mu$. To test the sensitivity of our results to the model
parameters we consider the cases $-\langle\bar q q
\rangle^{1/3}_{\rm ch} = 230$~MeV and $260$~MeV. The corresponding
parametrizations will be denoted as P230ch and P260ch,
respectively. For P230ch we find $\Lambda=608.3$~MeV and
$G\Lambda^2=28.43$, while for P260ch the parameter values are
found to be $\Lambda=914.6$~MeV and $G\Lambda^2=17.64$.

Given a set of parameters one can numerically solve the gap
equation (\ref{gapeq}) for different values of the chemical
potential and the external magnetic field. As mentioned above, for
given values of $\mu$ and $eB$ this equation has in general more
than one solution. In the chiral limit considered in this
subsection the solution $\bar \sigma=0$ is always present, while
for a fixed value of $eB$ and low enough values of $\mu$ a second
solution with a nonvanishing value $\bar\sigma$ also exists. In
particular, for $\mu=0$ the latter corresponds to the absolute
minimum of the thermodynamic potential, implying that the system
lies in a phase in which chiral symmetry is spontaneously broken.
Now, if one keeps $eB$ fixed and, starting from $\mu=0$, increases
the chemical potential, the value of the thermodynamic potential
corresonding to this solution remains unchanged, while that
corresponding to the trivial solution $\bar \sigma=0$ is found to
decrease, approaching the former. At some critical chemical
potential $\mu_c$ both values coincide, and for $\mu > \mu_c$ the
trivial solution is the one that corresponds to absolute minimum
of $\Omega_{\mu,B}^{\rm MFA,reg}$. Thus, at the critical value
$\mu = \mu_c$, which is in general a function of the magnetic
field, the system undergoes a transition to a phase in which
chiral symmetry is restored.
\begin{figure}[htb]
\centering
\includegraphics[width=0.47\textwidth]{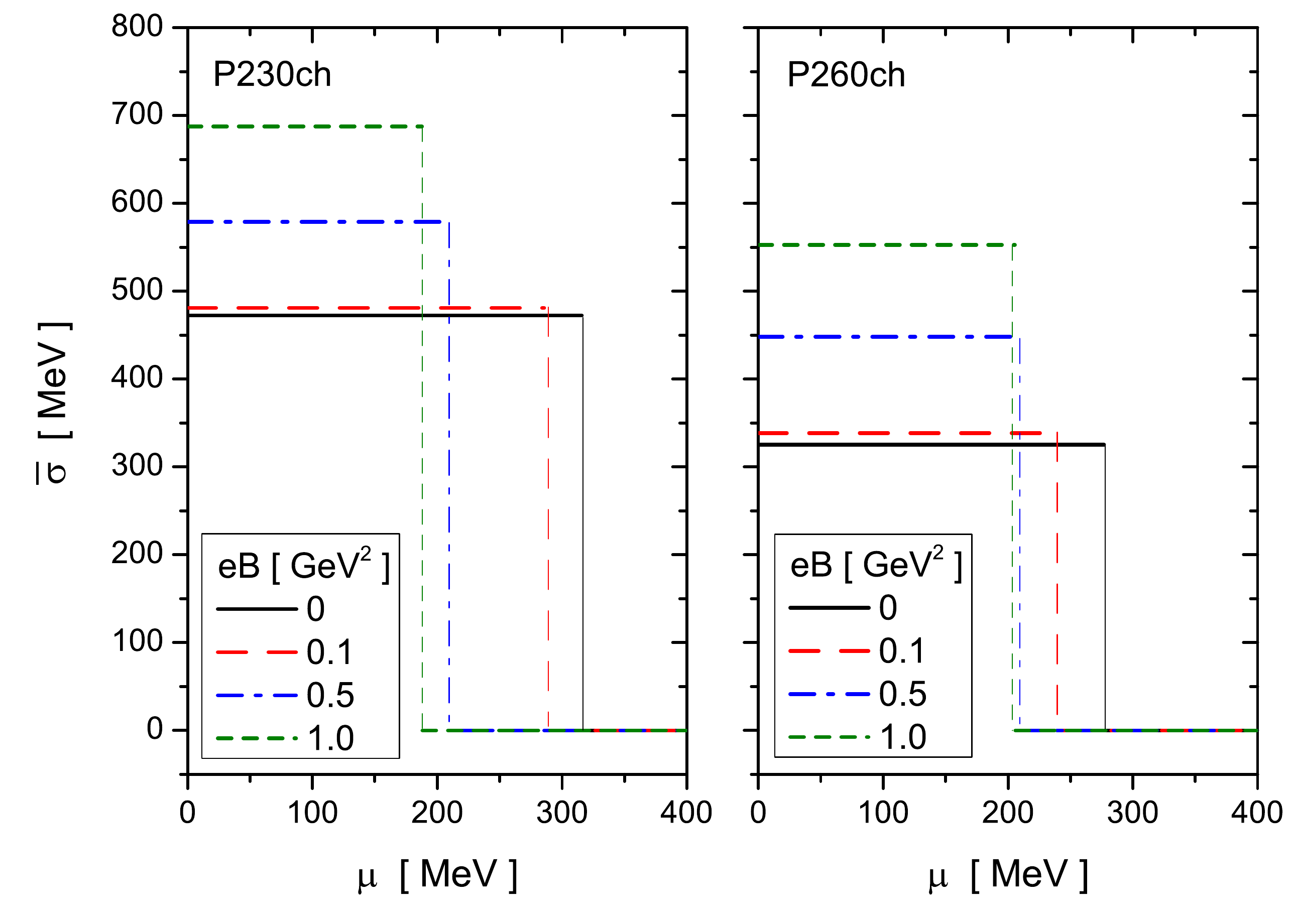}
\caption{Mean field value $\bar{\sigma}$ as a function of $\mu$
for some selected values of the $eB$, for parametrizations P230ch
(left) and P260ch (right).} \label{figura1}
\end{figure}

The behavior of $\bar \sigma$ as a function of $\mu$ for some
selected values of the magnetic field, and our two
pa\-ram\-e\-triza\-tions, is shown in Fig.~\ref{figura1}. The
vertical lines correspond to the critical chemical potentials
$\mu_c$, at which a first order transition is clearly observed. It
can be seen that for both parametrizations the value of $\bar
\sigma$ at $\mu=0$ gets enhanced if $eB$ is increased. This is a
manifestation of the well-known ``magnetic catalysis'' effect,
which entails a growth of the absolute value of the condensate
with $eB$ in vacuum. In fact, as shown in Refs.~\cite{Pagura:2016pwr,GomezDumm:2017iex}, the rate
of this increase is consistent with the results obtained through
lQCD simulations~\cite{Bali:2011qj,Bali:2012zg}.

\begin{figure}[htb]
\centering
\includegraphics[width=0.47\textwidth]{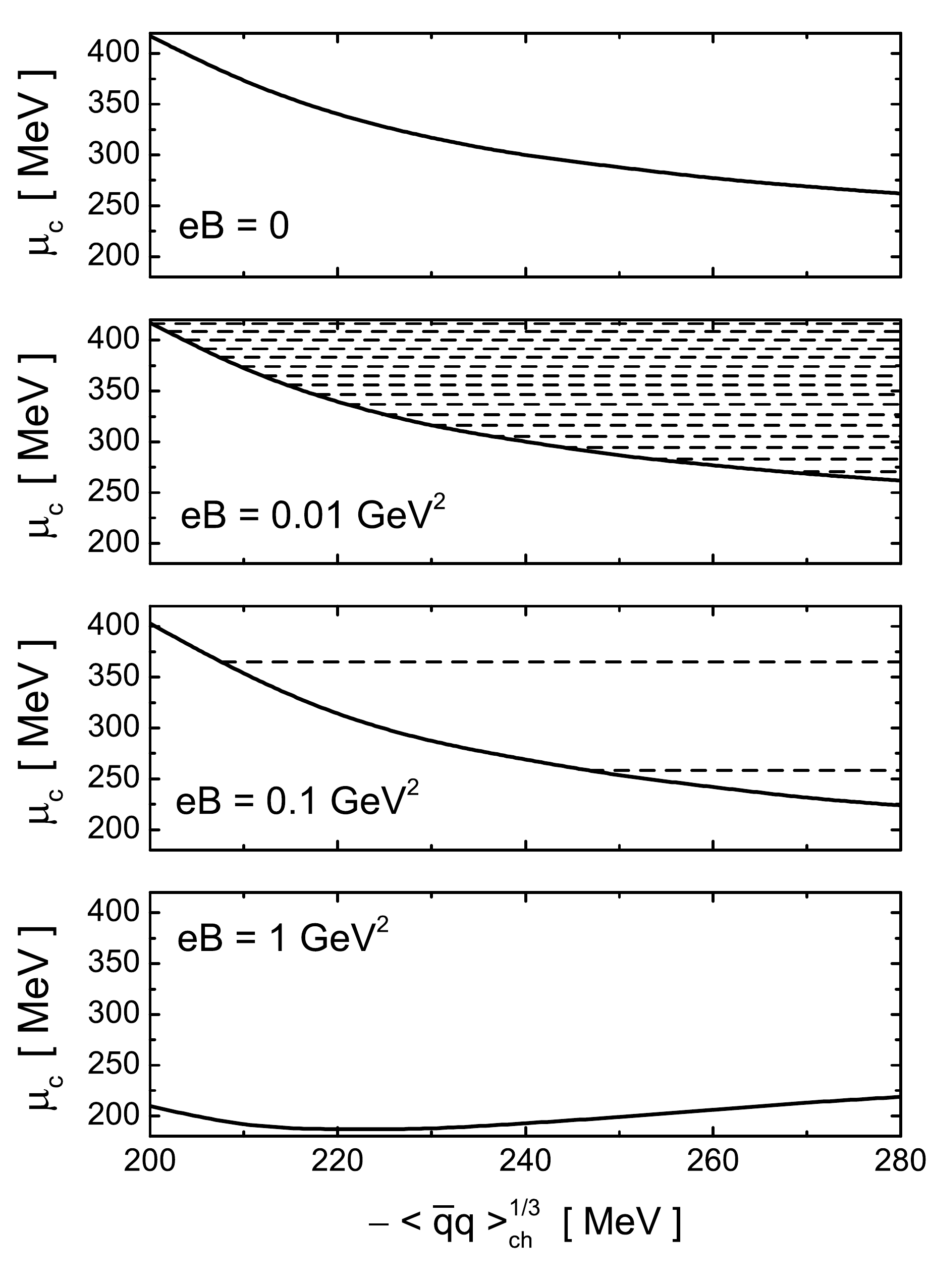}
\caption{Critical chemical potential as a function of the
parametrization choice, characterized by the value of the quark
chiral condensate at zero chemical potential and magnetic field.
The graphs correspond to parametrization P230ch and some
representative values of $eB$. The dashed lines indicate the
chemical potentials corresponding to the vAdH second order phase
transitions.} \label{fig2}
\end{figure}

It can also be observed that the parametrization choice has some
impact on the values of $\mu_c$ as well as on their dependence on
the magnetic field. To analyze this issue in more detail, in
Fig.~\ref{fig2} we plot the value of the critical chemical
potential as a function of the parametrization choice, for some
selected values of the magnetic field.
The parametrization is characterized by the value of $-\langle\bar
q q\rangle^{1/3}_{\rm ch}$ in the horizontal axis, which
corresponds to a given set of values of $\Lambda$ and $G$ (the
remaining input quantity is, as stated, the value of the pion
decay constant, $f_{\pi,{\rm ch}}=90$ MeV). Comparing the results
obtained for $eB=0$ and $eB=0.01$~GeV$^2$ (upper panels in
Fig.~\ref{fig2}) it is seen that although the full line ---that
corresponds to the first order phase transition between the chiral
symmetry broken phase and the chiral symmetry restored phase--- is
almost identical in both cases, for $eB=0.01$ the $\bar\sigma = 0$
region is subdivided into many phases. Following the notation in
Ref.~\cite{Ebert:1999ht}, we denote as phase \textbf{B} the chiral
symmetry broken phase and as $\textbf{A}_k$ ($k=0,1,2,\dots$) the
chiral symmetry restored phases that show up for finite $eB$. Each
one of the latter corresponds to a different number of populated
Landau levels, indicated by the index $k$.

The passage from any of the $\textbf{A}_k$ phases to the next one
is known as a van Alphen-de Haas (vAdH) transition. In the chiral
limit discussed in this subsection they are regulated by the
Heaviside theta function that appears in the last term of
Eq.~(\ref{omegafree}). Hence, the transition form the phase
$\textbf{A}_{k-1}$ to the phase $\textbf{A}_k$ happens at a
critical chemical potential given by
\begin{eqnarray}
\mu_{c,{\rm ch}}^{\rm vAdH} \ = \ \sqrt{ 2 k B_f}\ . \label{vAdH}
\end{eqnarray}
Clearly, this relation is independent of the parametrization,
which explains the fact that in Fig.~\ref{fig2} the dashed lines
associated to these transitions are parallel to the horizontal
axis. One should keep in mind that in the present case (i.e.\ for
$m_c=0$) there is no change in the order parameter when one goes
from one $\textbf{A}_k$ phase to the next one. In fact, one has
$\bar \sigma=0$ for all these phases, and, consequently, all these
transitions are of second order. The effect of vAdH transitions on
physical quantities can be seen, for example, in the quark density
$\rho_q = -\partial \Omega_{\mu,B}^{\rm MFA,reg}/\partial \mu$.
\begin{figure}[htb]
\centering
\includegraphics[width=0.47\textwidth]{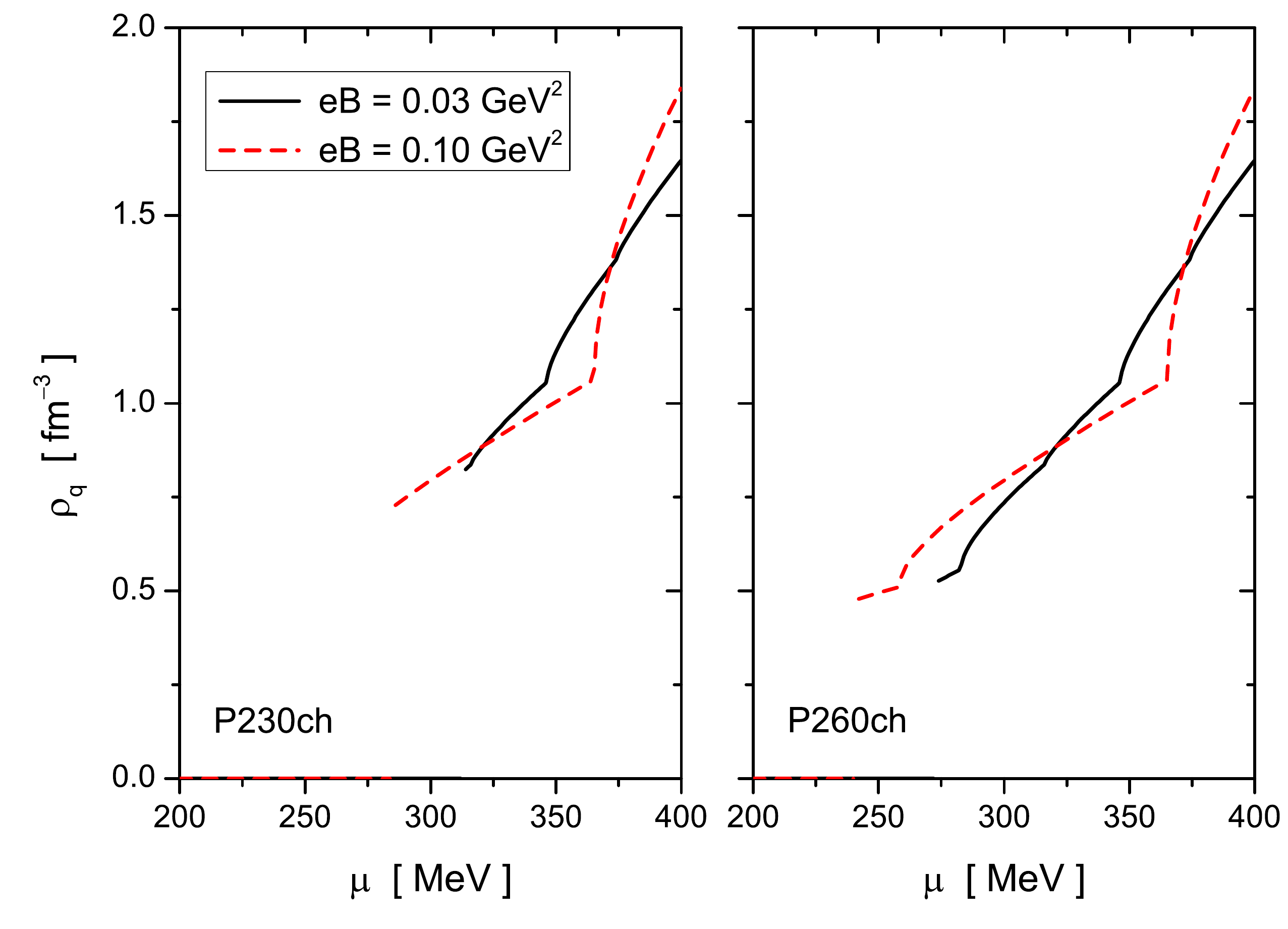}
\caption{Quark density as a function of the quark chemical
potential, for two representative values of $eB$ and
parametrizations P230ch (left) and P260ch (right).}
\label{density}
\end{figure}
This is illustrated in Fig.~\ref{density}, where we show the
behavior of $\rho_q$ as a function of $\mu$ for two representative
values of the magnetic field and both parametrizations P230ch and
P260ch. It can be observed that when the system moves from a given
phase $\textbf{A}_k$ to the next one the derivative of the density
shows a discontinuity at the transition point.

As it follows from Eq.~(\ref{vAdH}), the number of vAdH
transitions in a certain range of chemical potentials depends on
the magnetic field strength. This is the reason why in the panel
corresponding to $eB=0.01$~GeV$^2$ one observes a quite large
number of vAdH transitions, while for $eB=0.1$~GeV$^2$ there are
only two, and none is found for $eB=1$~GeV$^2$.

Another interesting point to be observed in Fig.~\ref{fig2} is
that the dependence of the critical chemical potential for the
first order transition with the parametrization (solid lines) is
quite similar for $eB=0$, 0.01~GeV$^2$ and 0.1~GeV$^2$. In all
these cases the value of $\mu_c$ decreases as the (absolute) value
of the condensate that characterizes the parametrization
increases. However, for $eB = 1$~GeV$^2$ the situation is
different. For parametrizations corresponding to $-\langle\bar q
q\rangle^{1/3}_{\rm ch} \gtrsim 220$~MeV we find that $\mu_c$
slightly increases with the absolute value of the condensate.
Thus, one can expect that for large values of $eB$ the behaviour
of $\mu_c$ would be more sensitive to the chosen parametrization.

In addition, it is important to remark that, for the whole range of
parametrizations considered in the pre\-sent nonlocal model, one
finds a direct transition from the phase {\bf B} to any of the
phases $\textbf{A}_k$. This differs from the situation observed
for the case of the local NJL model, in which for some
parametrizations one finds intermediate massive phases
$\textbf{C}_k$~\cite{Allen:2013lda}.

\begin{figure}[htb]
\includegraphics[width=0.47\textwidth]{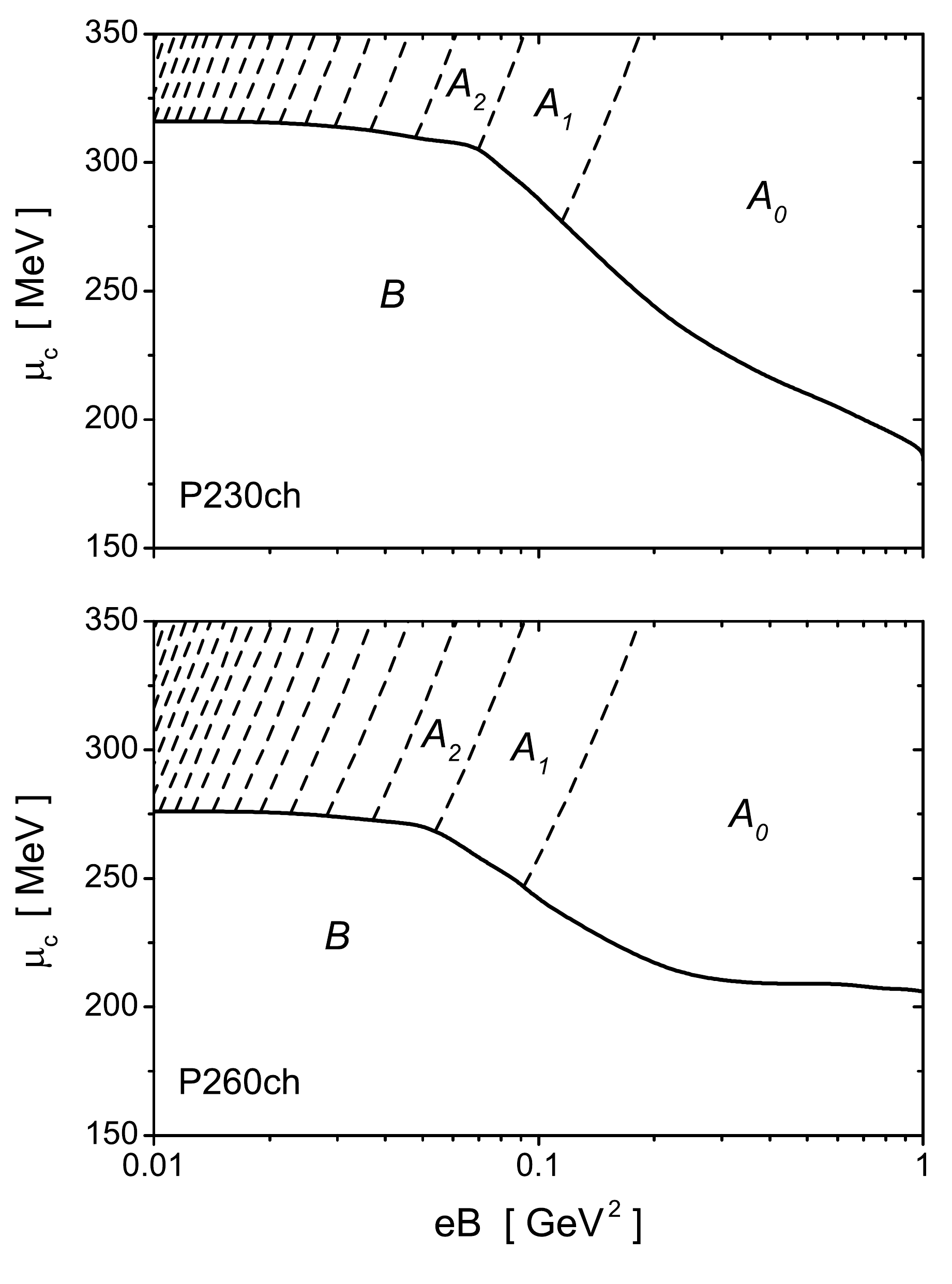}
\caption{Phase diagrams in the $eB-\mu$ plane in the chiral limit
case, for parametrizations P230ch (up) and P260ch (down). Solid
and dashed lines correspond to first and second order phase
transitions, respectively.} \label{PDch}
\end{figure}

We conclude this subsection by considering the phase diagrams of
magnetized cold quark matter in the $eB-\mu$ plane for
parametrizations P230ch and P260ch, shown in Fig.~\ref{PDch}.
As stated, \textbf{B} denotes the spontaneous chiral symmetry
broken phase, while $\textbf{A}_{k}$, $k=0,1,2,\dots$, correspond to chiral symmetry
restored phases. The second order transitions between
$\textbf{A}_{k}$ phases are indicated by the dashed lines. Each
time one of these lines is crossed from right to left, new Landau
levels get populated. The crossing from $\textbf{A}_0$ to
$\textbf{A}_1$ corresponds to the population of the $d$-quark
state with $k = 1$, while for the $u$-quark state only the lowest
level $k=0$ is allowed. Then, the crossing from $\textbf{A}_1$ to
$\textbf{A}_2$ implies the simultaneous population of the
$u$-quark state with $k = 1$ and the $d$-quark state with $k = 2$,
etc. The fact that the population of the up quark state with a
certain $k$ coincides with the one of the down quark state with
$2k$ is simply due to the fact that (in modulus) the electric
charge of the up quark is twice that of the down quark. Regarding
the behavior of the first order transition line (solid line in the
figure), we can observe three distinct regions. For $eB \lesssim
0.06$~GeV$^2$ the critical chemical potential $\mu_c$ depends only
weakly on $eB$, showing a slight decrease as the magnetic
field gets increased. Then, in the region from $eB \sim
0.06$~GeV$^2$ to $eB \sim 0.2$~GeV$^2$, a much pronounced decrease
of $\mu_c$ is observed. The behavior found in these two regions is
common to both parameterizations in Fig.~\ref{PDch}, and it has
also been observed in other models like the local NJL
model~\cite{Inagaki:2003yi,Allen:2013lda} and the Sakai-Sugimoto
model~\cite{Preis:2010cq}. This corresponds to the $\mu$IMC
effect mentioned in the Introduction. Now, the situation for magnetic fields larger
than $eB \sim 0.2$~GeV$^2$ turns out to depend on the chosen
parametrization, as it has been anticipated from the analysis of
Fig.~\ref{fig2}. In the case of the parametrization P230ch it is
observed that the decrease of $\mu_c$ continues with a rather
steep slope, whereas for P260ch it is found that the curve
$\mu_c(eB)$ becomes almost flat for $eB \gtrsim 0.2$~GeV$^2$. It
should be noted that for no reasonable parametrization of the
nonlocal NJL model we find a strong increase of $\mu_c$ with $eB$,
as it is observed in the local NJL model~\cite{Allen:2013lda}.
Actually, we have checked that a qualitatively similar behavior as
that found in Fig.~\ref{PDch} is also obtained for other nonlocal
form factor shapes, such as Lorenztian-like functions.

\subsection{Finite current quark mass}

In the case of a finite current quark mass, the input parameters
of the model are three, namely $\Lambda$, $G$ and $m_{c}$. They
can be set so as to reproduce the phenomenological values of the
pion mass and decay constant, $m_{\pi}=139$~MeV and
$f_{\pi}=92.4$~MeV, and some acceptable input value of the quark
condensate at zero $\mu$ and $B$. As in the chiral limit case, we
consider the values $-\left\langle \bar{q}q\right\rangle^{1/3} =
230$~MeV and 260~MeV. The corresponding parametrizations are
denoted as P230 and P260, respectively. For P230 one has $\Lambda
= 677.8$~MeV, $G \Lambda^2 = 23.65$ and $m_{c} = 6.4$~MeV, whereas
for P260 one has $\Lambda = 903.4$~MeV, $G \Lambda^2 = 17.53$ and
$m_{c} = 4.6$~MeV.

\begin{figure}[htb]
\includegraphics[width=0.47\textwidth]{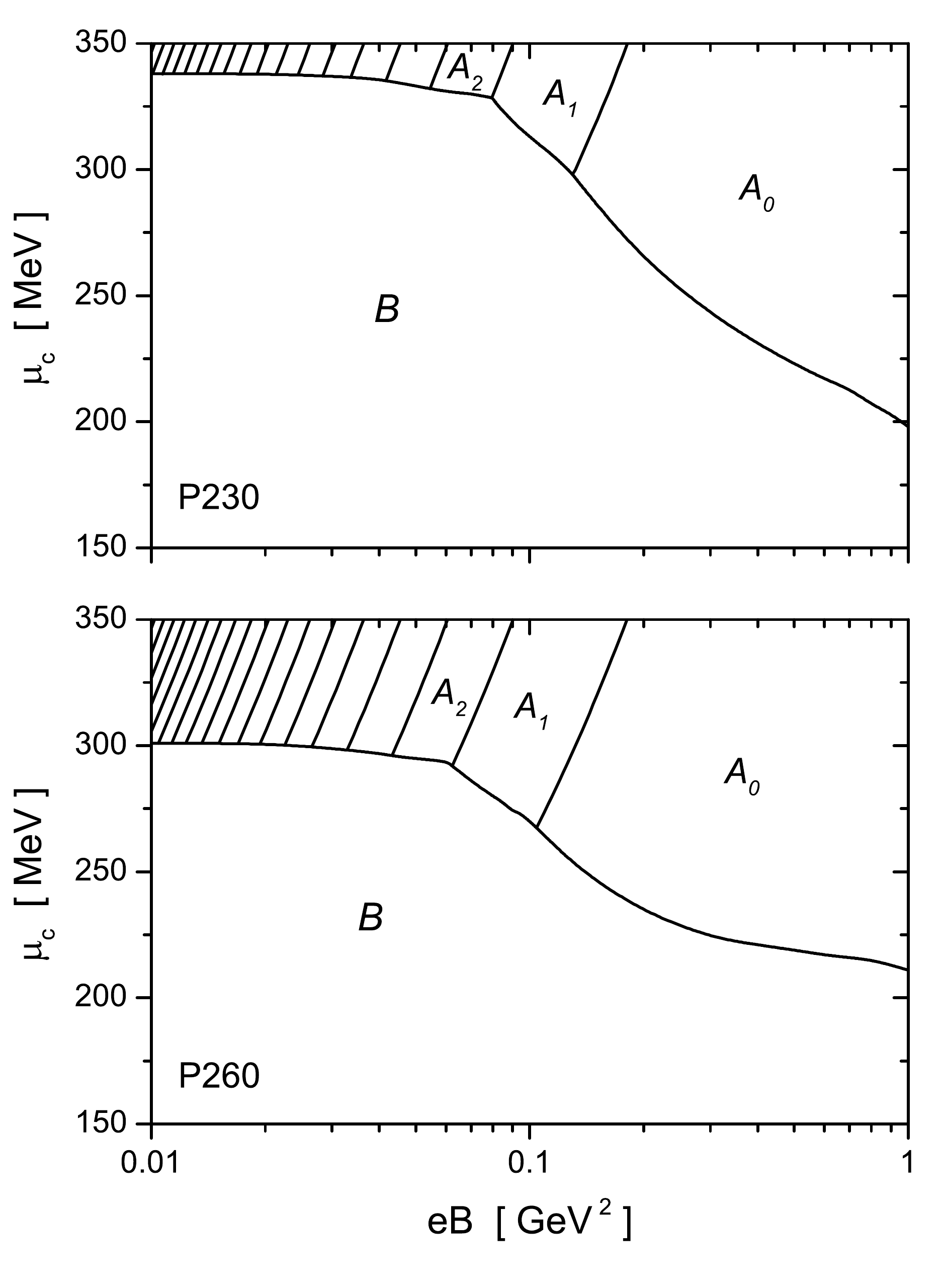}
\caption{Phase diagrams for the case of finite current quark
masses. Upper and lower panels correspond to parametrizations P230
and P260, respectively.} \label{fig5}
\end{figure}

As discussed in the previous subsection, once the input parameters
have been fixed one can numerically solve the gap equation
(\ref{gapeq}) for given values of the chemical potential and the
magnetic field strength. As in the chiral limit case, taking a
fixed value of $eB$, for $\mu=0$ the system always lies in a phase
$\textbf{B}$ in which chiral symmetry is spontaneously broken (the
absolute minimum of the thermodynamic potential occurs for a
solution with a relatively large value of $\bar\sigma$). Then, if
the chemical potential is increased, at some critical value $\mu =
\mu_c$ the system undergoes a transition to a phase
$\textbf{A}_k$, in which $\bar\sigma$ jumps to a small value,
indicating an approximate restoration of chiral symmetry. This is
shown in Fig.~\ref{fig5}, where we quote the phase diagrams in the
$eB-\mu$ plane corresponding to parametrizations P230 and P260.

\begin{figure}[htb]
\includegraphics[width=0.47\textwidth]{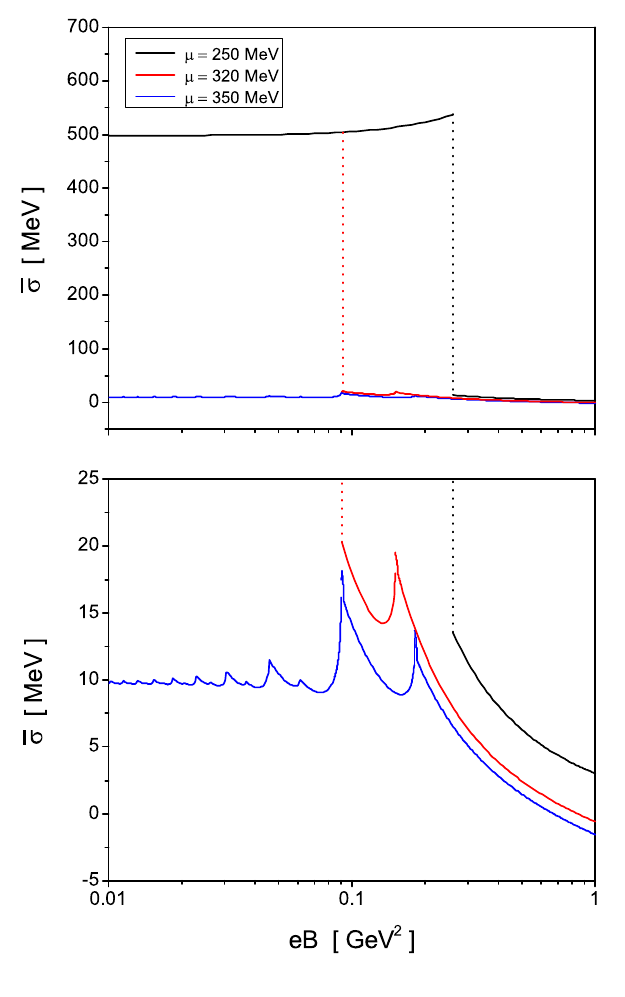}
\caption{Mean field value $\bar\sigma$ as a function of the magnetic field for
three representative values of the quark chemical potential. The
results correspond to parametrization P230.} \label{fig6}
\end{figure}

It can be seen that the phase structure is similar to that found
in the chiral limit case. Inverse magnetic catalysis is found for
finite $\mu$, as the critical chemical potential decreases for
increasing external magnetic field. In addition, it is found that
the phase space region in which chiral symmetry is approximately
restored is subdivided into many phases $\textbf{A}_{k}$,
$k=0,1,2,\dots$, which correspond to different population of the
Landau levels. However, notice that in this case the value of the
order parameter $\bar\sigma$ in the $\textbf{A}_{k}$ phases is
nonvanishing, therefore it can be used to signal not only the
chiral restoration transition but also the van Alphen-de Haas
transitions. This is illustrated in Fig.~\ref{fig6}, where we show
the behavior of $\bar\sigma$ as a function of $eB$ for
parametrization P230, taking three different fixed values of
$\mu$.
In the upper panel, first order chiral restoration transitions are
clearly seen for $\mu = 250$~MeV (black) and $\mu = 320$~MeV
(red), from phase $\textbf{B}$ to phases $\textbf{A}_0$ and
$\textbf{A}_1$, respectively. For $\mu = 350$~MeV the system lies
in the chiral restored region for all values of $eB$, therefore
the value of $\bar\sigma$ remains close to zero (blue). Now, the
vAdH transitions can be observed by looking at the lower panel of
Fig.~\ref{fig6}, in which we plot the same curves as in the upper
panel using a different scale and concentrating on the region of
small $\bar\sigma$. The curve corresponding to $\mu = 320$~MeV
(red) shows, beyond the chiral restoration transition from the
$\textbf{B}$ phase to the $\textbf{A}_1$ phase, a sharp peak
denoting a vAdH transition from the $\textbf{A}_1$ phase to the
$\textbf{A}_0$ one. A closer look indicates that there is a small
discontinuity in the left hand side of this peak, indicating that
the vAdH transition is in fact a first order one. The curve
corresponding to $\mu = 350$~MeV (blue) involves transitions
between many $\textbf{A}_k$ states, as can be seen from the phase
diagram in Fig.~\ref{fig5}. Each transition is characterized by
the presence of a peak in $\bar\sigma(eB)$, the height of these
peaks becoming smaller as the magnetic field gets decreased. As
far as numerical calculations can show, these vAdH transition are
of first order, therefore they are represented by solid lines in
Fig.~\ref{fig5}. As discussed for the chiral limit case, the vAdH
transitions correspond alternatively to the population of either a
new level for the $d$-quark state or a new level for both the $u$
and $d$-quark states. Therefore, they can be grouped into pairs,
leading to the pattern observed in the lower panel of
Fig.~\ref{fig6}. It is worth noticing that the location of the
vAdH transition curves in the $eB-\mu$ phase diagram shows just a
slight shift with respect to those found in the chiral limit case,
given by Eq.~(\ref{vAdH}).

\section{Conclusions}

In the present work we have studied the behavior of cold and dense
quark matter in the presence of an external homogeneous magnetic
field. We have considered a two-flavor nonlocal NJL model, in
which quark-antiquark currents include a Gaussian form factor.

It is found that for low values of the quark chemical potential
$\mu$ the system lies in a phase $\textbf{B}$ in which chiral
symmetry is sponteneously broken, while at some critical value
$\mu_c$ there is a first order phase transition in which this
symmetry becomes approximately restored. The restored phase can be
subdivided into many phases $\textbf{A}_k$, characterized by the
number of populated Landau levels for $u$ and $d$-quark states.
One important observation is that for the whole range of
parametrizations considered in the pre\-sent nonlocal model one
finds a direct transition from the phase {\bf B} to any of the
phases $\textbf{A}_k$. This differs from the situation observed
for the case of the local NJL model, in which for some
parametrizations one finds intermediate massive phases
$\textbf{C}_k$~\cite{Allen:2013lda}.

In the chiral limit $m_c =0$ it is seen that the van Alphen-de
Haas transitions between $\textbf{A}_k$ phases are of second
order, and their effect shows up e.g.~in the behaviour of the
quark density. On the contrary, for finite quark masses these
transitions are found to be of first order, though the
corresponding jumps in the order parameter $\bar\sigma$ are rather
tiny. Concerning the first order chiral restoration transition
line, it is found that up to $eB \sim 0.2$~GeV$^2$ the critical
chemical potential $\mu_c$ decreases with the magnetic field,
showing an inverse magnetic catalysis effect. For larger values of
$eB$ the behaviour of $\mu_c$ becomes more or less flat, depending
on the parametrization. In any case, for the considered
parametrization range we do not find a significant growth of the
critical chemical potential for large magnetic fields, as occurs
in the case of the local NJL model.

\begin{acknowledgements}
This work has been supported in part by Consejo
Nacional de Investigaciones Cient\'ificas y T\'ecnicas and Agencia Nacional
de Promoci\'on Cient\'ifica y Tecnol\'ogica (Argentina), under Grants
No.~PIP17-700 and No.~PICT17-03-0571, respectively, and by the National
University of La Plata (Argentina), Project No.~X824.
\end{acknowledgements}

%
%
%


\begin{thebibliography}{199}

\bibitem{HIC}
D.~E.~Kharzeev, L.~D.~McLerran and H.~J.~Warringa, Nucl.\ Phys.\ A
{\bf 803}, 227 (2008); V. Skokov, A. Y. Illarionov, and V. Toneev,
Int. J. Mod. Phys. A {\bf 24}, 5925 (2009); V. Voronyuk, V.
Toneev, W. Cassing, E. Bratkovskaya, V. Konchakovski, and S.
Voloshin, Phys. Rev. C {\bf 83}, 054911 (2011).

\bibitem{cosmo}
T. Vachaspati, Phys. Lett. {\bf B265}, 258 (1991); K. Enqvist and
P. Olesen, Phys. Lett. {\bf B319}, 178 (1993).

\bibitem{duncan}
R. C. Duncan and C. Thompson, Astrophys. J. 392, L9 (1992); C.
Kouveliotou et al., Nature {\bf 393}, 235 (1998).

\bibitem{Kharzeev:2012ph}
D.~E.~Kharzeev, K.~Landsteiner, A.~Schmitt and H.~U.~Yee,
Lect.\ Notes Phys.\  {\bf 871}, 1 (2013).

\bibitem{Andersen:2014xxa}
  J.~O.~Andersen, W.~R.~Naylor and A.~Tranberg,
  Rev.\ Mod.\ Phys.\  {\bf 88}, 025001 (2016).

\bibitem{Miransky:2015ava}
  V.~A.~Miransky and I.~A.~Shovkovy,
  Phys.\ Rept.\  {\bf 576}, 1 (2015).

\bibitem{Nambu:1961tp}
Y.~Nambu and G.~Jona-Lasinio,
Phys. Rev. \textbf{122}, 345-358 (1961)

\bibitem{Nambu:1961fr}
Y.~Nambu and G.~Jona-Lasinio,
Phys. Rev. \textbf{124}, 246-254 (1961)


\bibitem{Vogl:1991qt}
U.~Vogl and W.~Weise,
Prog. Part. Nucl. Phys. \textbf{27}, 195-272 (1991)

\bibitem{Klevansky:1992qe}
S.~P.~Klevansky,
Rev. Mod. Phys. \textbf{64}, 649-708 (1992)

\bibitem{Hatsuda:1994pi}
T.~Hatsuda and T.~Kunihiro,
Phys. Rept. \textbf{247}, 221-367 (1994)
[arXiv:hep-ph/9401310 [hep-ph]]

\bibitem{Schmidt:1994di}
S.~M.~Schmidt, D.~Blaschke and Y.~L.~Kalinovsky,
Phys. Rev. C \textbf{50}, 435-446 (1994)

\bibitem{Burden:1996nh}
C.~J.~Burden, L.~Qian, C.~D.~Roberts, P.~C.~Tandy and
M.~J.~Thomson,
Phys. Rev. C \textbf{55}, 2649-2664 (1997)
[arXiv:nucl-th/9605027 [nucl-th]]

\bibitem{Bowler:1994ir}
R.~D.~Bowler and M.~C.~Birse,
Nucl. Phys. A \textbf{582}, 655-664 (1995)
[arXiv:hep-ph/9407336 [hep-ph]]


\bibitem{Ripka:1997zb}
G.~Ripka, ``Quarks bound by chiral fields: The quark-structure of
the vacuum and of light mesons and baryons,'' (Oxford University
Press, Oxford, 1997)

\bibitem{Dumm:2021vop}
D.~G.~Dumm, J.~P.~Carlomagno and N.~N.~Scoccola,
Symmetry \textbf{13}, no.1, 121 (2021)
 [arXiv:2101.09574
[hep-ph]].

\bibitem{Gusynin:1994re}
V.~P.~Gusynin, V.~A.~Miransky and I.~A.~Shovkovy,
Phys. Rev. Lett. \textbf{73}, 3499-3502 (1994) [erratum: Phys.
Rev. Lett. \textbf{76}, 1005 (1996)]


\bibitem{Bali:2011qj}
  G.~S.~Bali, F.~Bruckmann, G.~Endrodi, Z.~Fodor, S.~D.~Katz, S.~Krieg, A.~Schafer and K.~K.~Szabo,
  JHEP {\bf 1202}, 044 (2012).

\bibitem{Bali:2012zg}
  G.~S.~Bali, F.~Bruckmann, G.~Endrodi, Z.~Fodor, S.~D.~Katz and A.~Schafer,
  Phys.\ Rev.\ D {\bf 86}, 071502 (2012).


\bibitem{Pagura:2016pwr}
V.~P.~Pagura, D.~Gomez Dumm, S.~Noguera and N.~N.~Scoccola,
Phys. Rev. D \textbf{95}, 034013 (2017)
[arXiv:1609.02025 [hep-ph]]

\bibitem{GomezDumm:2017iex}
D.~Gomez Dumm, M.~F.~Izzo Villafa\~ne, S.~Noguera, V.~P.~Pagura
and N.~N.~Scoccola,
Phys. Rev. D \textbf{96}, 114012 (2017)
[arXiv:1709.04742 [hep-ph]]

\bibitem{GomezDumm:2017jij}
D.~Gomez Dumm, M.~F.~Izzo Villafa\~ne and N.~N.~Scoccola,
Phys. Rev. D \textbf{97},  034025 (2018)
[arXiv:1710.08950 [hep-ph]]

\bibitem{Dumm:2020muy}
D.~Gomez Dumm, M.~F.~Izzo Villafa\~ne and N.~N.~Scoccola,
Phys. Rev. D \textbf{101},  116018 (2020)
[arXiv:2004.10052 [hep-ph]]



\bibitem{Ebert:1999ht}
  D.~Ebert, K.~G.~Klimenko, M.~A.~Vdovichenko and A.~S.~Vshivtsev,
  Phys.\ Rev.\ D {\bf 61}, 025005 (2000)
  [hep-ph/9905253].

\bibitem{Ebert:2003yk}
  D.~Ebert and K.~G.~Klimenko,
  Nucl.\ Phys.\ A {\bf 728}, 203 (2003)
  [hep-ph/0305149].

\bibitem{Inagaki:2003yi}
  T.~Inagaki, D.~Kimura and T.~Murata,
  Prog.\ Theor.\ Phys.\  {\bf 111}, 371 (2004)
  [hep-ph/0312005].

\bibitem{Menezes:2008qt}
  D.~P.~Menezes, M.~Benghi Pinto, S.~S.~Avancini, A.~Perez Martinez and C.~Providencia,
  Phys.\ Rev.\ C {\bf 79}, 035807 (2009)
  [arXiv:0811.3361 [nucl-th]].

\bibitem{Boomsma:2009yk}
  J.~K.~Boomsma and D.~Boer,
  Phys.\ Rev.\ D {\bf 81}, 074005 (2010)
  [arXiv:0911.2164 [hep-ph]].



\bibitem{Ferrari:2012yw}
  G.~N.~Ferrari, A.~F.~Garcia and M.~B.~Pinto,
  Phys.\ Rev.\ D {\bf 86}, 096005 (2012)
  [arXiv:1207.3714 [hep-ph]].

\bibitem{Ferrer:2012wa}
  E.~J.~Ferrer and V.~de la Incera,
  Lect.\ Notes Phys.\  {\bf 871}, 399 (2013)
  [arXiv:1208.5179 [nucl-th]].

\bibitem{Allen:2013lda}
P.~G.~Allen and N.~N.~Scoccola,
Phys. Rev. D \textbf{88}, 094005 (2013)

\bibitem{Avancini:2012ee}
S.~S.~Avancini, D.~P.~Menezes, M.~B.~Pinto and C.~Providencia,
Phys. Rev. D \textbf{85}, 091901 (2012)
[arXiv:1202.5641 [hep-ph]].

\bibitem{Grunfeld:2014qfa}
A.~G.~Grunfeld, D.~P.~Menezes, M.~B.~Pinto and N.~N.~Scoccola,
Phys. Rev. D \textbf{90}, no.4, 044024 (2014)
[arXiv:1402.4731 [hep-ph]].

\bibitem{Allen:2015qxa}
P.~G.~Allen, V.~P.~Pagura and N.~N.~Scoccola,
Phys. Rev. D \textbf{91}, no.11, 114024 (2015)
[arXiv:1502.00572 [hep-ph]].


\bibitem{Preis:2010cq}
  F.~Preis, A.~Rebhan and A.~Schmitt,
  JHEP {\bf 1103}, 033 (2011)
  [arXiv:1012.4785 [hep-th]];

\bibitem{Preis:2012fh}
  F.~Preis, A.~Rebhan and A.~Schmitt,
  Lect.\ Notes Phys.\  {\bf 871}, 51 (2013)
  [arXiv:1208.0536 [hep-ph]].


\bibitem{Fayazbakhsh:2010gc}
  S.~Fayazbakhsh and N.~Sadooghi,
  Phys.\ Rev.\ D {\bf 82}, 045010 (2010)
  [arXiv:1005.5022 [hep-ph]].

\bibitem{Mandal:2012fq}
T.~Mandal and P.~Jaikumar,
Phys. Rev. C \textbf{87}, 045208 (2013)
[arXiv:1209.2432 [nucl-th]].

\bibitem{Allen:2015paa}
P.~G.~Allen, A.~G.~Grunfeld and N.~N.~Scoccola,
Phys. Rev. D \textbf{92}, no.7, 074041 (2015)
[arXiv:1508.04724 [hep-ph]].


\bibitem{Mandal:2017ihr}
T.~Mandal and P.~Jaikumar,
Adv. High Energy Phys. \textbf{2017}, 6472909 (2017)
[arXiv:1701.02561 [hep-ph]].

\bibitem{Coppola:2017edn}
M.~Coppola, P.~Allen, A.~G.~Grunfeld and N.~N.~Scoccola,
Phys. Rev. D \textbf{96}, no.5, 056013 (2017)
[arXiv:1707.03795 [hep-ph]].

\bibitem{Bloch:1952qkt}
C.~Bloch,
Kong. Dan. Vid. Sel. Mat. Fys. Med. \textbf{27N8}, no.8, 1-55 (1952)

\bibitem{Ritus:1978cj}
V.~I.~Ritus,
Sov.\ Phys.\ JETP {\bf 48}, 788 (1978)

\bibitem{GomezDumm:2004sr}
D.~Gomez Dumm and N.~N.~Scoccola,
Phys. Rev. C \textbf{72}, 014909 (2005)
[arXiv:hep-ph/0410262 [hep-ph]]

\end{thebibliography}

\end{document}